\documentclass[11pt,a4paper,english,nofootinbib,,superscriptaddress]{revtex4}
\usepackage{lmodern}

\usepackage[T1]{fontenc}
\usepackage[latin9]{inputenc}
\usepackage{babel}

\usepackage{amsmath}
\usepackage{graphicx}
\usepackage{amssymb}
\usepackage{esint}
\usepackage[unicode=true, pdfusetitle,
 bookmarks=true,bookmarksnumbered=false,bookmarksopen=false,
 breaklinks=false,pdfborder={0 0 1},backref=false,colorlinks=false]
 {hyperref}
\def\b{\begin{equation}}
\def\e{\end{equation}}

\makeatletter
\@ifundefined{textcolor}{}
{%
 \definecolor{BLACK}{gray}{0}
 \definecolor{WHITE}{gray}{1}
 \definecolor{RED}{rgb}{1,0,0}
 \definecolor{GREEN}{rgb}{0,1,0}
 \definecolor{BLUE}{rgb}{0,0,1}
 \definecolor{CYAN}{cmyk}{1,0,0,0}
 \definecolor{MAGENTA}{cmyk}{0,1,0,0}
 \definecolor{YELLOW}{cmyk}{0,0,1,0}
 }

\usepackage{latexsym}\usepackage{bm}

\makeatother

\makeatother

\begin{document}

\title{Weyl-gauging of Topologically Massive Gravity}

\author{Suat Dengiz}

\email{suat.dengiz@metu.edu.tr}

\affiliation{Department of Physics,\\
 Middle East Technical University, 06800, Ankara, Turkey}

\author{Ercan Kilicarslan}

\email{kercan@metu.edu.tr}

\affiliation{Department of Physics,\\
 Middle East Technical University, 06800, Ankara, Turkey}

\author{Bayram Tekin}

\email{btekin@metu.edu.tr}

\affiliation{Department of Physics,\\
 Middle East Technical University, 06800, Ankara, Turkey}

\date{\today}
\begin{abstract}

We construct a Weyl-invariant extension of topologically massive gravity
which, remarkably,  turns out to include topologically massive electrodynamics, with a Proca mass term, conformally coupled to a scalar field. The action has no dimensionful parameters, therefore,  the masses are generated via symmetry breaking either radiatively in flat backgrounds or spontaneously in constant curvature backgrounds. The broken phase of the theory, generically, has a single massive spin-2 and a massive spin-1 excitation. Chiral gravity in asymptotically anti-de Sitter spacetimes  does not arise as a low energy theory, while chiral gravity in de Sitter spacetime is not ruled out.
\end{abstract}
\maketitle

\section{Introduction}

Unlike its lower spin cousins, interacting spin-2 theories in $ 3+1 $ dimensions suffer from two seemingly unrelated problems: The first one being the non-existence of a proper quantum field theory and the second 
one being the non-existence of a covariant mass. Clearly, the first problem is about the high energy regime while the second is about the low energy, in fact classical, regime of the theory.
What is interesting is that to get a better UV behaved theory one usually adds more powers of curvature to the Einstein-Hilbert action as
\begin{equation}
{\cal L}=\sigma R+\alpha R^2+ \beta R^2_{\mu \nu},
\label{hordc}
\end{equation}
which then necessarily has a massive spin-2 particle in its spectrum in addition to the massless spin-2 (and massless spin-0) particles \cite{stelle}. One might be tempted to conclude that the above two problems could be related, but since (\ref{hordc}) is renormalizable but non-unitary, such a conclusion is not warranted. A similar situation exists in $2+1$ dimensions for the particular combination of parameters $8 \alpha + 3 \beta =0$, which is known as the  new massive gravity (NMG) \cite{BHT}: The theory has a massive spin-2 unitary excitation but fails to be renormalizable \cite{muneyuki}.

On the other hand, in $ 2+1 $ dimensions, there is another dynamical theory of gravity, the topologically massive gravity (TMG)\cite{DJT} which is both renormalizable \cite{DZ} and unitary that propagates a massive spin-2 particle (with a single helicity mode). What is rather intriguing is that TMG with a tuned gravitational Chern-Simons parameters in terms of the (negative) cosmological constant in an asymptotically $ AdS_3 $ spacetime  might actually be a consistent quantum gravity theory albeit a chiral one \cite{Li:2008dq} with a dual chiral $2D$ conformal field theory (CFT) on the boundary. Hence TMG is a candidate for which the above mentioned two problems might be related.  A natural question about TMG is  the following: Can one extend the theory in such a way that mass of the spin-2 particle arises as a result of a symmetry breaking as in the case of Higgs-mechanism for lower spin particles? In \cite{DengizTekin, Tanhayi:2011aa, Tanhayi:2012nn} such a mechanism was shown to exist for  NMG and its infinite curvature 
extension Born-Infeld-NMG \cite{binmg}.  Here we will answer the question in the affirmative by finding  a Weyl-invariant version of TMG which  necessarily  incorporates topologically massive electrodynamics (TME) and a Proca mass term.  The action is scale-free but  symmetry breaking takes place either spantaneously in the  $(A)dS$ vacuum or at two loop level about the flat vacuum.   
Spin-2 and  spin-1 particles  become massive as a result of the symmetry breaking. We find the perturbative spectrum of the theory and study its tree level unitarity by expanding the action up to quadratic order in the fluctuations of the fields.

The layout of the paper is as follows: In section II, we construct the Weyl-gauging of TMG and find the masses of the fundamental excitations and also study the fluctuations about the vacuum of the theory. Section III is devoted to the TME-Proca theory in $(A)dS$. We conclude with Section IV and give the field equations in the Appendix.

\section{Weyl-gauging of TMG}

It is well-known that Einstein's gravity in 2+1 dimensions, with or without a cosmological constant, does not have any bulk propagating degrees of freedom. This situation does not change if the theory is conformally coupled to a scalar field as  
\begin{equation}
 S=\int d^3 x \sqrt{-g} \Big (\Phi^2 R +8 \partial_\mu \Phi \partial^\mu \Phi-\frac{\nu \Phi^6}{2} \Big).
\label{cieha}
\end{equation}
Even though it is not apparent from this action as it seems to have at least a ghost\footnote{We work with the mostly plus signature.} scalar particle, this is a red herring as one can show \cite{Tanhayi:2011aa} that conformally coupled scalar-tensor theory still does not have any propagating degrees of freedom by either directly expanding the action
around its $dS$ or flat vacuum (for $ \nu=0 $) up to quadratic order in the fields or by going directly to the Einstein frame with the transformation  $ g_{\mu \nu}(x)=(\frac{\Phi}{\Phi_0})^2 g^{E}_{\mu\nu}(x)$. Note that the requirement that one can go to the Einstein frame from the Jordan frame introduces a
 dimensionful scale $ \Phi^2_0 $ which is the inverse of the Newton's constant. The latter method  transforms (\ref{cieha}) to pure cosmological Einstein gravity
\begin{equation}
 S=\int d^3x \sqrt{-g^E} \,\, \Phi^2_0 \Big (R^E-\frac{\nu}{2} \Phi^4_0 \Big ).
\end{equation}
On the other hand, if one adds the third derivative order, parity non-invariant, gravitational Chern-Simons term to the Einstein-Hilbert action, one gets the TMG
\begin{equation}
 S_{TMG}=\int d^3 x \sqrt{-g} \bigg [ \sigma m R +\frac{k}{2 } \epsilon^{\lambda \mu \nu} \Big ( \Gamma^\rho_{\lambda \sigma} \partial_\mu \Gamma^\sigma_{\nu \rho}+\frac{2}{3} \Gamma^\rho_{\lambda \sigma} \Gamma^\sigma_{\mu \tau}\Gamma^\tau_{\nu \rho}  \Big ) \bigg ],
\label{tmg}
\end{equation}
with a single spin-2 propagating degree of freedom with mass $ M_{graviton}=-\frac{\sigma m}{\lvert k \rvert} $ around its flat vacuum. Here $ \epsilon^{\lambda \mu \nu} $ is a \emph{rank-3 tensor} and $ \sigma,k $ are dimensionless and also $ m>0$ is of mass dimension, for unitarity $\sigma <0$ must be chosen in flat space.
If a cosmological constant is added to (\ref{tmg}) as $ -2 \Lambda$, then the mass of the spin-2 excitation is shifted as $M^2_{graviton}=\frac{\sigma^2 m^2}{k^2 }+\Lambda $ \cite{Carlip:2008jk, Gurses:2011fv}. Furthermore, if  the cosmological constant is tuned as 
\begin{equation}
 \Lambda=-\frac{\sigma^2 m^2}{k^2},
\end{equation}
one obtains the so called chiral gravity \cite{Li:2008dq} with no bulk degrees of freedom (save the log-modes \cite{grum})  but a chiral boundary conformal theory. [See 
\cite{Bagchi:2012yk} for a recent discussion of how chiral gravity might appear in flat spaces.]

Chern-Simons term is invariant under the diffeomorphisms and conformal scalings of the metric only up to a boundary term in both cases. Obviously, TMG is not invariant under conformal transformations. One can
take $ \sigma=0 $ leaving the pure gravitational Chern-Simons theory which was studied before as a pure gauge theory \cite{Horne:1988jf} or more more recently in \cite{Afshar:2011qw} in the context of holography. 
Another route is to use (\ref{cieha}) to obtain a conformally invariant version of TMG (again up to a boundary term) which reads \cite{Deser:2004wd}
\begin{equation}
\begin{aligned}
 S_{CTMG}=\int d^3 x \sqrt{-g} \bigg [&\sigma \Phi^2 R +8 \partial_\mu \Phi \partial^\mu \Phi-\frac{\nu \Phi^6}{2}\\
& + \frac{k}{2 } \epsilon^{\lambda \mu \nu} \Big ( \Gamma^\rho_{\lambda \sigma} \partial_\mu \Gamma^\sigma_{\nu \rho}+\frac{2}{3} \Gamma^\rho_{\lambda \sigma} \Gamma^\sigma_{\mu \tau} \Gamma^\tau_{\nu \rho}  \Big ) \bigg ].  
\label{ctmg}
\end{aligned}
\end{equation}
It is clear that if the scalar field has a non-zero vacuum expectation value $ \langle \Phi \rangle=m^{1/2}  $, then (\ref{ctmg}) reproduces TMG (\ref{tmg}). Note that as discussed in \cite{DengizTekin,Tanhayi:2011aa,Tanhayi:2012nn},
for the flat backgrounds the scalar field will necessarily assume a vacuum expectation value due to radiative corrections at two-loop level \cite{tantekin} as in the Coleman-Weinberg mechanism in $3+1$ dimensions \cite{coleman}.
On the other hand for constant curvature spaces, conformal symmetry will be broken spontaneously in the vacuum. While this procedure provides a nice symmetry breaking origin to the topological mass in $2+1$ dimensions,
in this paper, we shall be interested in a more general symmetry and its broken phase: We will show that Weyl-gauging of TMG (\ref{tmg}) will yield an action which unifies TMG and its abelian gauge field cousin TME \cite{DJT} whose action is 

\begin{equation}
S_{TME}=\int d^3x \sqrt{-g} \Big [ -\frac{1}{4}F_{\mu\nu}^2+\frac{\mu}{4} \epsilon^{ \mu \nu \lambda} F_{\mu\nu} A_\lambda \Big ], 
\end{equation}
 and a Proca mass term. In flat backgrounds, TME has a single spin-1 excitation with mass $ M_{gauge}=\lvert \mu \rvert $. TME-Proca theory has two spin-1 helicity modes with different masses. (See how these masses get shifted in $(A)dS$ below). As discussed at length in \cite{ioro, DengizTekin,Tanhayi:2011aa,Tanhayi:2012nn}, Weyl-gauging of a Poincar\'e invariant theory
is equivalent to upgrading rigid scale invariance, $ x^\mu \rightarrow \lambda x^\mu $ and $ \Phi \rightarrow \lambda^d \Phi $ where $ d $ is the scaling dimension of the field, to a local one. This is  implemented by introducing a gauge field (Weyl-gauge field)
$ A^\mu $ together with the following transformations of the fields and the metric, while keeping $ x^\mu $ intact,
\begin{equation}
\begin{aligned}
 &g_{\mu\nu} \rightarrow g^{'}_{\mu\nu}=e^{2 \zeta(x)} g_{\mu\nu}, \hskip 1 cm \Phi \rightarrow \Phi^{'} =e^{-\frac{\zeta(x)}{2}} \Phi, \\
&{\cal{D}}_\mu \Phi =\partial_\mu\Phi -\frac{1}{2} A_\mu \Phi, \hskip 1 cm {\cal{D}}_\mu g_{\alpha \beta}=\partial_\mu g_{\alpha\beta}+ 2 A_\mu g_{\alpha \beta}, \\
&A_\mu \rightarrow A^{'}_\mu = A_\mu - \partial_\mu \zeta(x),
\label{transform1}
\end{aligned}
\end{equation}
where we denoted gauge covariant derivative with $ {\cal{D}}_\mu $ to distinguish it from the spacetime covariant derivative $ \nabla_\mu $ to appear below. Note also that, we have written our formulas in $2+1$ dimensions.

To find the Weyl-gauged version of (\ref{tmg}), we can use the following Weyl-invariant Christoffel connection
\begin{equation}
 \widetilde{\Gamma}^\lambda_{\mu\nu}=\frac{1}{2}g^{\lambda\sigma} \Big ({\cal{D}}_\mu g_{\sigma\nu}+{\cal{D}}_\nu g_{\mu\sigma}
-{\cal{D}}_\sigma g_{\mu\nu} \Big),
\label{christofel}
\end{equation}
or explicitly
\begin{equation}
 \widetilde{\Gamma}^\lambda_{\mu \nu} =\Gamma^\lambda_{\mu \nu}+\delta^\lambda_\nu A_\mu+\delta^\lambda_\mu A_\nu-g_{\mu \nu} A^\lambda,
\label{gammm1}
\end{equation}
which yields a Weyl-invariant Riemann tensor as
 \begin{equation}
\widetilde{R}^\mu{_{\nu\rho\sigma}} [g,A]
=R^\mu{_{\nu\rho\sigma}}+\delta^\mu{_\nu}F_{\rho\sigma}+2
\delta^\mu_{[\sigma} \nabla_{\rho]} A_\nu
+2 g_{\nu[\rho}\nabla_{\sigma]} A^\mu \\
 +2 A_{[\sigma} \delta_{\rho]}^\mu A_\nu +2 g_{\nu[\sigma}
A_{\rho]} A^\mu +2 g_{\nu[\rho} \delta_{\sigma]}^\mu  A^2,
\label{wiriem}
\end{equation}
and a Weyl-invariant Ricci tensor
\begin{equation}
\begin{aligned}
\widetilde{R}_{\nu\sigma} [g,A]&= \widetilde{R}^\mu{_{\nu\mu\sigma}}[g,A] \\
&=R_{\nu\sigma}+F_{\nu\sigma}-(n-2)\Big [\nabla_\sigma A_\nu - A_\nu A_\sigma +A^2  g_{\nu\sigma} \Big ]-g_{\nu\sigma}\nabla \cdot A,
\label{wiricc}
\end{aligned}
\end{equation}
where $\nabla\cdot A\equiv \nabla_\mu A^\mu$. One more contraction gives the scalar curvature
\begin{equation}
\widetilde{R}[g,A]=R-2(n-1)\nabla \cdot A-(n-1)(n-2) A^2,
\label{wisclcr}
\end{equation}
which is {\it {not}} Weyl-invariant, but, can be made Weyl-invariant with a compensating scalar field.

Collecting all the pieces, Weyl-gauged version of TMG (\ref{tmg}) can be written as

\begin{equation}
 \begin{aligned}
S_{WTMG}=&\int d^3x  \sqrt{-g}\,\, \sigma \Phi^2 [R-4 \nabla . A-2 A^2] \\
&+ \frac{k}{2}\int d^3 x \sqrt{-g} \,\, \epsilon^{\lambda \mu \nu} \Big ( \tilde{\Gamma}^\rho_{\lambda \sigma} \partial_\mu \tilde{\Gamma}^\sigma_{\nu \rho}+\frac{2}{3} \tilde{\Gamma}^\rho_{\lambda \sigma} \tilde{\Gamma}^\sigma_{\mu \tau} \tilde{\Gamma}^\tau_{\nu \rho}  \bigg ). 
\label{witmg1} 
\end{aligned}
\end{equation}
Denoting the Lagrangian density of the gravitational Chern-Simons part of (\ref{witmg1}) as $ CS(\widetilde{\Gamma}) $, one can show, with the help of (\ref{gammm1}), that
\begin{equation}
 \begin{aligned}
  CS(\widetilde{\Gamma})=CS(\Gamma)+ \frac{k}{4}\epsilon^{\lambda \mu \nu} A_\lambda F_{\mu\nu}  -\partial_\mu \Big [ \frac{k}{2}\epsilon^{\lambda \mu \nu} g^{\alpha \sigma} (\partial_\lambda g_{\nu \sigma}) A_\alpha- \frac{k}{2} \epsilon^{\lambda \mu \nu} \Gamma^\rho_{\lambda \rho} A_\nu  \Big ].
 \end{aligned}
\end{equation}
Therefore up to a boundary term, Weyl-invariant TMG action reads
\begin{equation}
 \begin{aligned}
S_{WTMG}&=\int d^3x  \sqrt{-g} \,\,\sigma \Phi^2 [R-4 \nabla . A-2 A^2] \\
&+ \frac{k}{2}\int d^3 x \sqrt{-g} \,\, \epsilon^{\lambda \mu \nu} \bigg ( \Gamma^\rho_{\lambda \sigma} \partial_\mu \Gamma^\sigma_{\nu \rho}+\frac{2}{3} \Gamma^\rho_{\lambda \sigma} \Gamma^\sigma_{\mu \tau} \Gamma^\tau_{\nu \rho}  \bigg )
+ \frac{k}{4}\int d^3 x \sqrt{-g} \,\, \epsilon^{\lambda \mu \nu} A_\lambda F_{\mu \nu}.  
\label{witmgac} 
\end{aligned}
\end{equation}
It is important to note that Weyl invariance is more general than the conformal invariance. If one takes the Weyl gauge to be a pure-gauge as
\begin{equation}
 A_\mu=2\partial_\mu \ln \Phi,
\end{equation}
then Weyl-invariant TMG (\ref{witmgac}) reproduces the conformally-invariant TMG (\ref{ctmg}) (up to the scalar potential which can be added by hand as was done in(\ref{ctmg})).
The fact that abelian Chern-Simons term comes from the Weyl-gauging of gravitational Chern-Simons term is quite interesting. In some sense Weyl-gauging unifies gravitational and abelian Chern-Simons theories.
Needless to say that abelian Chern-Simons theory does not arise in the conformally-invariant version of TMG: it only arises in the Weyl-invariant version.

We can also add the usual Weyl-invariant scalar matter part and the Weyl-invariant non-minimally coupled Maxwell part 
\begin{equation}
 S_{\Phi}=-\frac{\alpha}{2} \int d^3x \sqrt{-g}\,\,(D_\mu \Phi D^\mu \Phi+\nu \Phi^6) \,, \hskip 0.5cm \hskip 1cm S_{A^\mu}=-\frac{\beta}{4}\int d^3x \sqrt{-g}\,\, \Phi^{-2} F_{\mu\nu}F^{\mu \nu}.
\end{equation}
Here all the parameters $ \sigma,k,\alpha,\beta $ are dimensionless, as imposed by Weyl invariance. One of the parameters could be set to $1$ but we keep them to have the freedom to set any of them to zero. Mass dimensions of the fields are as: 
\begin{equation}
 [g_{\mu\nu}]=M^0=1 \hskip 1cm ; \hskip 1cm [\Phi]=M^{1/2} \hskip 1cm ; \hskip 1cm [A_\mu]=M.
\end{equation}
Finally, let us collect all the pieces and write the Lagrangian density of the Weyl-invariant TMG:
\begin{equation}
 \begin{aligned}
{\cal L}_{WTMG}&=\sigma \Phi^2 [R-4 \nabla . A-2 A^2]+ \frac{k}{2} \epsilon^{\lambda \mu \nu} \bigg ( \Gamma^\rho_{\lambda \sigma} \partial_\mu \Gamma^\sigma_{\nu \rho}+\frac{2}{3} \Gamma^\rho_{\lambda \sigma} \Gamma^\sigma_{\mu \tau} \Gamma^\tau_{\nu \rho}  \bigg )\\
&+ \frac{k}{4} \epsilon^{\lambda \mu \nu} A_\lambda F_{\mu\nu} -\frac{\alpha}{2}(D_\mu \Phi D^\mu \Phi+\nu \Phi^6)-\frac{\beta}{4}\Phi^{-2} F_{\mu\nu}F^{\mu \nu}.
\label{witmg2} 
\end{aligned}
\end{equation}
In the symmetric vacuum, $  \langle \Phi \rangle=0 $, the only term that survives is the gravitational Chern-Simons term without a propagating degree of freedom. The theory is conformally-invariant and the Weyl gauge field has to vanish due to the Maxwell term. [It is clear that the symmetric vacuum is a singular point.  In fact whenever higher curvature terms either for the gauge field  (Maxwell term here) or the gravity part, such as  $R^2$, are introduced in the Weyl-invariant setting, the symmetric and the non-symmetric vacua are disconnected.  This is not possible to avoid with higher curvature terms. But of course with such an action, one could argue that symmetry is necessarily broken and the  $\langle \Phi \rangle=0$  is not allowed.] On the other hand in the broken phase with  $  \langle \Phi \rangle=m^{1/2} $, (\ref{witmg2}) becomes
\begin{equation}
 \begin{aligned}
{\cal L}_{WTMG}&= \sigma m R-\frac{\alpha \nu}{2}m^3+ \frac{k}{2} \epsilon^{\lambda \mu \nu} \bigg ( \Gamma^\rho_{\lambda \sigma} \partial_\mu \Gamma^\sigma_{\nu \rho}+\frac{2}{3} \Gamma^\rho_{\lambda \sigma} \Gamma^\sigma_{\mu \tau} \Gamma^\tau_{\nu \rho}  \bigg ) \\
&-\frac{\beta}{4m} F_{\mu\nu}F^{\mu \nu}+ \frac{k}{4} \epsilon^{\lambda \mu \nu} A_\lambda F_{\mu\nu}-\frac{m}{2} \Big (4 \sigma +\frac{\alpha}{4} \Big ) A^2.
\label{vwitmg} 
\end{aligned}
\end{equation}
The first line is the cosmological TMG and the second line is the TME-Proca theory  coupled to gravity. Relying on earlier works  \cite{Carlip:2008jk, Gurses:2011fv}, we can easily read the mass of the graviton in the $(A)dS$ vacuum:
\begin{equation}
 M^2_{graviton}=\frac{m^2 \sigma^2}{k^2}+\Lambda \hskip 0.5cm \mbox{where} \hskip 0.5cm \Lambda=\frac{\alpha \nu m^2}{4 \sigma}.
\label{grav_mass}
\end{equation}
As for the TME-Proca sector, all we know from the Literature is the masses of the two helicity modes in flat backgrounds  \cite{tantekin, Dunne:1998qy} given as 
\begin{equation}
 M^\pm_{gauge}(\Lambda=0)= \frac{1}{2} \bigg \{\sqrt{\frac{k^2m^2}{\beta^2}+\frac{4m}{\beta} \Big (4 \sigma+\frac{\alpha}{4}\Big)}\pm \frac{m \lvert k \rvert}{\beta}  \bigg \}.
\label{mass_gauge_3}
\end{equation}
In the special case of vanishing Proca term, that is when $ 16 \sigma+\alpha=0 $, only one of the helicity-1 mode survives with a mass $ M_{+}= \frac{m \lvert k \rvert}{\beta} $, the other mode becomes a pure gauge (this does not follow from  (\ref{mass_gauge_3}), one should go back to (\ref {vwitmg}) to see it ). 
In the next section, we will find the mass of the gauge field in TME-Proca theory in $(A)dS$, but here let us quote the final result
\begin{equation}
M^2_{gauge_\pm}(\Lambda\ne0)=\frac{15 \Lambda}{4}+M^2_{gauge_\pm}(\Lambda = 0).
\label{gauge_kutle}
\end{equation}
Let us now discuss the tree-level unitarity of the theory, that is its tachyon and ghost-freedom, both in flat and $(A)dS$ backgrounds. In flat backgrounds unitarity requires
\begin{equation}
 \sigma <0, \hskip 1cm \beta>0 \hskip 0.5cm \mbox{and} \hskip 0.5cm \alpha+\frac{k^2 m}{\beta} \ge -16 \sigma.
\end{equation}
In AdS backgrounds ($ \Lambda<0 $), we have more possibilities. For spin-2 particles Breitenlohner-Freedmann (BF) bound \cite{bf,waldron} should be satisfied, $ M^2_{graviton} \ge \Lambda $, which is satisfied in our case.
For the gauge field we should have $ M^2 _{gauge} (\Lambda) \ge 0$ which leads to a bound on $ \Lambda $ as
\begin{equation}
 \Lambda \ge -\frac{4}{15} M^2_{gauge}(\Lambda=0).
\end{equation}
For dS backgrounds ($ \Lambda>0 $) Higuchi bound \cite{higuchi} $ M^2_{graviton}\ge\Lambda>0 $ should be satisfied which does not bring any condition except the existence of a dS vacuum
requires $ \sigma>0 $ (assuming $\alpha>0, \nu>0$).

In studying the perturbative spectrum of the Weyl-invariant TMG above, we have frozen the scalar field to its vacuum value. One could suspect that this procedure cannot be conclusive in exploring the unitarity and the stability
of the theory since it does not take into account the fluctuations in the direction of the scalar field. But as the following explicit computation reveals, the scalar mode is actually non-dynamical:
If $ \Phi $ is not zero, by a choice of gauge it can be made to be constant. Let us now prove this assertion by naively expanding the action (\ref{witmg2}) up to quadratic order in all the fields as was done in \cite{Gullu:2010em,Tanhayi:2011aa,Tanhayi:2012nn}
\begin{equation}
\begin{aligned}
 &\Phi\equiv \sqrt{m}+\tau \Phi^L,\hskip 0.8cm g_{\mu\nu} \equiv \bar{g}_{\mu\nu}+\tau h_{\mu\nu}, \hskip 0.8cm A_\mu \equiv \tau A^L_\mu, \\
& g^{\mu\nu}=\bar{g}^{\mu\nu}-\tau h^{\mu\nu}+\tau^2 h^{\mu\rho}h^\nu_\rho, \hskip 0.8cm \sqrt{-g}=\sqrt{-\bar{g}}\,[1+\frac{\tau}{2} h+\frac{\tau^2}{8}(h^2-2h^2_{\mu\nu})], \\
&\nabla_\mu A_\alpha =\tau \bar{\nabla}_\mu A^L_\alpha-\tau^2 (\Gamma^\gamma_{\mu\alpha})_L A^L_\gamma-\tau^2 h^\gamma_\beta (\Gamma^\beta_{\mu\alpha})_L A^L_\gamma, 
\end{aligned}
\end{equation}
where we have introduced a small dimensionless parameter $ \tau $ to keep track of the order of expansions. Using these results, and the vacuum equation $ \Lambda=\frac{\alpha \nu m^2}{4 \sigma} $ which follows from the $ {\cal O}(\tau) $ expansion of the action or from the full non-linear field equations given in the Appendix, (\ref{witmg2}) at $ {\cal O}(\tau^2) $, becomes
\begin{equation}
\begin{aligned}
 {I}_{WTMG}^{(2)}=\int d^3x\sqrt{-\bar{g}}\bigg \{& -\frac{\alpha}{2}(\partial_\mu \Phi^L)^2 -6 \alpha \nu m^2 \Phi^2_L-\sqrt{m} \Big ( 8 \sigma +\frac{\alpha}{2} \Big ) \Phi^L \bar{\nabla}\cdot A^L \\
& -\frac{\beta}{4m}(F^L_{\mu\nu})^2+\frac{k}{4} \epsilon^{\lambda \mu \nu} A^L_\lambda F^L_{\mu\nu}- m \Big ( 2 \sigma +\frac{\alpha}{8} \Big ) A^2_L \\
&-\frac{\sigma m}{2} h^{\mu \nu} {\cal G}^{L}_{\mu \nu}+\frac{k}{2}h^{\mu \nu} C^{L}_{\mu \nu} + 2 \sigma \sqrt{m} \Phi^L R^L \bigg \},
\label{linearform}
\end{aligned}
\end{equation}
up to irrelevant boundary terms which we dropped. Here the linearized tensors are \cite{deser_tekin_en}
\begin{equation}
\begin{aligned}
 C_{L}^{\mu \nu}&=\frac{\epsilon^{\mu \alpha\beta}}{\sqrt{-\bar{g}}} \bar{g}_{\beta \sigma} \nabla_\alpha \Big ( R^{\sigma\nu}_L-2 \Lambda h^{\sigma \nu}- \frac{1}{4} \bar{g}_{\beta \nu}  R_L \Big ),\hskip 0.7cm {\cal G}_{\mu\nu}^L=R_{\mu\nu}^L-\frac{1}{2}\bar{g}_{\mu\nu}R^L-2\Lambda h_{\mu\nu}, \\
R^{L}_{\nu \sigma}&=\frac{1}{2} \Big (\bar{\nabla}_\mu \bar{\nabla}_\sigma h^\mu_\nu+\bar{\nabla}_\mu \bar{\nabla}_\nu
 h^\mu_\sigma- \bar{\Box}h_{\sigma \nu}-\bar{\nabla}_\sigma \bar{\nabla}_\nu h \Big), \hskip 0.5cm R^{L}=\bar{\nabla}_\mu \bar{\nabla}_\nu h^{\mu \nu}-\bar{\Box}h-2 \Lambda h.
\end{aligned}
\end{equation}
We still have to decouple the terms in (\ref{linearform}), to this end redefining the perturbations as
\begin{equation}
h_{\mu\nu} \equiv\widetilde{h}_{\mu\nu}-\frac{4}{\sqrt{m}}\bar{g}_{\mu\nu}\Phi_L \hskip 1cm \mbox{and} \hskip 1cm A^L \equiv \widetilde{A}_\mu+\frac{2}{\sqrt{m}} \partial_\mu \Phi_L ,
\label{rede}
\end{equation}
removes the coupling between the fields
\begin{equation}
\begin{aligned}
 {I}_{WTMG}^{(2)}=\int d^3x\sqrt{-\bar{g}}\bigg \{& -\frac{\beta}{4m}(\widetilde{F}^L_{\mu\nu})^2+\frac{k}{4} \epsilon^{\lambda \mu \nu} \widetilde{A}^L_\lambda \widetilde{F}^L_{\mu\nu} - m \Big ( 2 \sigma +\frac{\alpha}{8} \Big ) \widetilde{A}^2_L \\
& -\frac{\sigma m}{2} \widetilde{h}^{\mu \nu} \Big [ \widetilde{{\cal G}^{L}}_{\mu \nu}- \frac{k}{\sigma m} \widetilde{C}^{L}_{\mu \nu} \Big ] \bigg \}, 
\label{linearform1}
\end{aligned}
\end{equation}
where the relevant shifted tensors read
\begin{equation}
 \begin{aligned}
(R_{\mu\nu})_L=&(\widetilde{R}_{\mu\nu})_L+\frac{2}{\sqrt{m}}(\bar{\nabla}_\mu\partial_\nu\Phi_L+\bar{g}_{\mu\nu}\bar{\Box}\Phi_L),
\hskip 0.7cm  R_L=\widetilde{R}_L+\frac{8}{\sqrt{m}}(\bar{\Box}\Phi_L+3\Lambda\Phi_L),\\
  {\cal G}_{\mu\nu}^L=&\widetilde{{\cal
G}}^L_{\mu\nu}+\frac{2}{\sqrt{m}}\Big(\bar{\nabla}_\mu\partial_\nu\Phi_L-\bar{g}_{\mu\nu}\bar{\Box}\Phi_L-2\Lambda
\bar{g}_{\mu\nu}\Phi_L\Big), \hskip 0.7cm \widetilde{h}^{\mu \nu} \widetilde{C}^{L}_{\mu \nu}=h^{\mu \nu} C^{L}_{\mu \nu},\\
h^{\mu\nu}{\cal G}^L_{\mu\nu}=&\widetilde{h}^{\mu\nu}{\cal
\widetilde{G}}^L_{\mu\nu}+\frac{4}{\sqrt{m}}\widetilde{R}_L\Phi_L+\frac{16}{m}\Phi_L\bar{\Box}\Phi_L+\frac{48}{m}\Lambda\Phi_L^2.
 \end{aligned}
\end{equation}
As expected  from the discussion of the previous section, (\ref{linearform1}) describes a parity-non-invariant  massive single spin-2 and a parity-non-invariant massive spin-1 excitation
whose masses are given respectively as  (\ref{grav_mass}) and (\ref{gauge_kutle}).  Let us now show how the latter mass can be computed. 

\section{Topologically Massive Electrodynamics-Proca theory  in $ (A)dS$}

Consider the field equation of the TME-Proca theory in a generic background (we shall specify to $(A)dS$ at the end )
\begin{equation}
 a \nabla_\nu F^{\nu \mu}+b \epsilon^{\lambda \nu \mu} F_{\lambda \nu}+c A^\mu=0,
\label{mcspe1}
\end{equation}
with $ a\ne0$ ; $ b\ne0$ ; $ c\ne0$ . We would like to find the masses of the fundamental excitations, so we must convert  (\ref{mcspe1}) to a wave type equation. Taking the divergence of (\ref{mcspe1}), one gets
\begin{equation}
\nabla_\mu A^\mu=0.
\end{equation}
Namely, for $c \ne 0$,  Lorenz gauge is imposed, which removes 1 out of the 3 possible dynamical degrees of freedom. [ Note that if $c=0$, then Lorenz gauge can be chosen.] Defining 
\begin{equation}
 \widetilde{F}^\mu=\frac{1}{2} \epsilon^{\mu \lambda \nu } F_{\lambda \nu} , \hskip 1cm {\cal B}^\mu=\epsilon^{ \mu \lambda \nu} \nabla_\lambda \widetilde{F}_\nu,
\end{equation}
 it is easy to show that $ {\cal B}^\mu=\nabla_\alpha F^{\alpha \mu} $.  Applying the operator  $\epsilon^{\alpha \nu \mu}  \nabla_\nu $ to (\ref{mcspe1} ) yields
\begin{equation}
 a \epsilon^{\alpha \nu \mu} \nabla_\nu {\cal B}_\mu+2b\epsilon^{\alpha \nu \mu}\nabla_\nu \widetilde{F}_\mu+c\epsilon^{\alpha \nu \mu}\nabla_\nu A_\mu=0.
\label{mcspe3}
\end{equation}
With the help of 
\begin{equation}
 \epsilon^{\alpha \nu \mu} \nabla_\nu {\cal B}_\mu=\Box \widetilde{F}^\alpha-R^\alpha{_\beta} \widetilde{F}^\beta,
\label{mcspe5}
\end{equation}
one can get
\begin{equation}
 {\cal B}^\alpha=-\frac{1}{2 b} \Big[a (\Box \widetilde{F}^\alpha-R^\alpha{_\beta} \widetilde{F}^\beta)+ c \widetilde{F}^\alpha \Big].
\label{bdenklemi}
\end{equation}
Applying  $\epsilon^{\sigma \lambda}{_\alpha}  \nabla_\lambda $ to  (\ref{mcspe3}) and using $\nabla_\alpha {\cal B}^\alpha=0$, which follows from the Bianchi identity, 
one obtains 
\begin{equation}
a \Box {\cal B}^\sigma-a R^\sigma{_\alpha}{\cal B}^\alpha+2 b \epsilon^{\sigma \lambda \alpha} \nabla_\lambda {\cal B}_\alpha+c {\cal B}^\sigma=0.
\label{mcspe7}
\end{equation}
Plugging (\ref{bdenklemi}) into  (\ref{mcspe7}) gives a fourth-order equation for TME-Proca theory in a generic background 
\begin{equation}
\begin{aligned}
 &\bigg [-\frac{a^2}{2b} \delta^\sigma_\beta \Box^2+ \Big (\frac{a^2}{b} R^\sigma{_\beta}-\frac{ac}{b} \delta^\sigma_\beta+2b \delta^\sigma_\beta \Big)\Box \\
&+\frac{a^2}{2b} (\Box  R^\sigma{_\beta} ) 
- \frac{a^2}{2b}R^\sigma{_\alpha} R^\alpha{_\beta}+\Big (\frac{ac}{b}-2b \Big) R^\sigma{_\beta}-\frac{c^2}{2b}\delta^\sigma_\beta \bigg] \widetilde{F}^\beta=0.
\label{mcspe8}
\end{aligned}
 \end{equation}
Specifically, by setting $ R^\alpha{_\beta}=2 \Lambda \delta^\alpha{_\beta} $, we get the corresponding result for  $(A)dS$ 
\begin{equation}
 \bigg [-\frac{a^2}{2b} \Box^2+\Big (\frac{2 \Lambda a^2}{b}-\frac{ac}{b}+2b \Big) \Box+\Big(-\frac{2a^2 \Lambda^2}{b}+\frac{2ac\Lambda}{b}-4b\Lambda-\frac{c^2}{2b} \Big) \bigg ] \widetilde{F}^\sigma=0.
\label{mcspe9}
\end{equation}
If one further sets $\Lambda =0$ one gets the flat space case 
\begin{equation}
 \bigg [ -\frac{a^2}{2b} \partial^4+ \Big(2b-\frac{ac}{b} \Big) \partial^2-\frac{c^2}{2b} \bigg]\widetilde{F}^\beta=0,
\end{equation}
from which the masses follow as 
\begin{equation}
 M^\pm_{gauge}(\Lambda=0)= \frac{1}{2} \bigg \{\sqrt{\frac{k^2m^2}{\beta^2}+\frac{4m}{\beta} \Big (4 \sigma+\frac{\alpha}{4}\Big)}\pm \frac{m \lvert k \rvert}{\beta}  \bigg \},
\end{equation}
where we have put  $ a=\frac{\beta}{m};b=\frac{k}{2};c=-\chi=-m (4\sigma+\alpha/4) $.

For $(A)dS$ the equation (\ref{mcspe9}) can be factored as
\begin{equation}
 \frac{\beta^2}{m^2}(\Box-\xi^2_+)(\Box-\xi^2_-) \widetilde{F}^\sigma=0,
\label{mcspe11}
\end{equation}
where
\begin{equation}
 \xi^2_\pm \equiv 2 \Lambda +M^2_{gauge_\pm}(\Lambda=0).
\end{equation}
From (\ref{mcspe11}) it is clear that there are \emph{two} propagating degrees of freedom with generically inequivalent
masses, as required in this parity non-invariant theory. To actually read the masses of these spin-1 degrees of freedom 
we have to recall that in $ D=2+1 $-dimensional $AdS$ space, a massless gauge field, that is a gauge field that propagates
on the null cone obeys \emph{not}$ \,$ $ \Box A^\mu=0 $ but
\begin{equation}
 \Big (\Box +\frac{7}{4} \Lambda \Big) A^\mu=0,
\end{equation}
where we assumed the Lorenz gauge $ \nabla_\mu A^\mu=0 $ \cite{Deser:1983mm}. Therefore from (\ref{mcspe11}) it follows that 
helicity $ \pm1 $ components of the gauge field have the  masses given by (\ref{gauge_kutle}).

\section{Conclusions}

In this work, we have shown that Weyl-gauging of the gravitational Chern-Simons term generates the abelian Chern-Simons term.  This result, augmented with the other Weyl-invariant scalar field-compensated Maxwell and Einstein actions, yields the most general Weyl-gauged topologically massive gravity coupled to topologically massive electrodynamics with a Proca term. We have calculated the particle spectrum and studied the quadratic fluctuations of the theory around its non-symmetric vacuum. Both the spin-2 and spin-1 excitations get their masses via symmetry breaking of the Weyl-symmetry. 

A natural question is whether or not chiral gravity arises in the broken phase, low energy limit of the theory. To answer this, we must first find a way to calculate conserved charges in this model. Fortunately, this was done in \cite{dtc} where it was shown that conserved charges defined in \cite{deser_tekin_en} are intact under conformal transformations (or change at most by a multiplicative constant) if the conformal scaling of the metric does not dramatically change the symmetry and the boundary structures. Relying on these results, we can show that the existence of an $AdS$ vacuum ( $\Lambda <0$ which requires  $\sigma <0$ )  is not consistent with the positivity of either left or right central charges. Therefore chiral gravity does not arise as a critical point in the Weyl-gauged TMG. On the other hand, $dS$ case  ( for $ \sigma >0$  ) still might be compatible with a conformal field theory in $S^2$ along the lines of 
\cite{strom_ds} which deserves a further investigation.
 
\section{\label{ackno} Acknowledgments}

The work of  B.T. is supported by the TUBITAK Grant No. 110T339. S.D. is supported by  TUBITAK Grant No. 109T748. E.K. is supported by the TUBITAK PhD Scholarship.

\section*{Appendix: Field Equations of WITMG}

We have found the maximally symmetric vacuum from the ${\cal{O}}(\tau)$ expansion of the action, but, it pays to check this result from the full non-linear equations which we give here. 
In any case, these equations are needed for future work on the exact solutions etc. of the theory. 
 
The variation of (\ref{witmg2}) with respect to $ g^{\mu \nu} $ results in 
\begin{equation}
\begin{aligned}
 &\sigma \bigg [ \Phi^2 G_{\mu \nu}+ g_{\mu \nu} \Box \Phi^2-\nabla_\mu \nabla_\nu \Phi^2-4\Phi^2 \nabla_\mu A_\nu +2g_{\mu \nu} \Phi^2\nabla . A -2\Phi^2 A_\mu A_\nu+g_{\mu \nu} \Phi^2 A^2 \bigg ] \\
+&\frac{\alpha}{4} g_{\mu \nu} D_\alpha \Phi  D^\alpha \Phi+\frac{\alpha \nu}{4} g_{\mu \nu} \Phi^6-\frac{\alpha}{2} D_\mu \Phi D_\nu \Phi+\frac{\beta}{8} g_{\mu \nu}\Phi^{-2} F^2_{\alpha \beta}+\frac{\beta}{2} \Phi^{-2} F_{\mu \alpha} F^\alpha{_\nu} +k C_{\mu \nu}=0.
\label{jkhl}
\end{aligned}
\end{equation}
With the help of $ D_\mu \Phi=\nabla_\mu \Phi-\frac{1}{2} A_\mu \Phi $, the equation (\ref{jkhl})  can be written as
\begin{equation}
\begin{aligned}
 & \sigma \Phi^2 G_{\mu \nu}+(\sigma-\frac{\alpha}{8}) g_{\mu \nu} \Box \Phi^2-(\sigma-\frac{\alpha}{4})\nabla_\mu \nabla_\nu \Phi^2-(4\sigma+\frac{\alpha}{4})\Phi^2 \nabla_\mu A_\nu +(2\sigma+\frac{\alpha}{8})g_{\mu \nu} \Phi^2\nabla . A \\
&-(2\sigma+\frac{\alpha}{8})\Phi^2 A_\mu A_\nu+(\sigma+\frac{\alpha}{16})g_{\mu \nu} \Phi^2 A^2 +\frac{\alpha}{4} g_{\mu \nu}(\nabla_\alpha \Phi)^2+\frac{\alpha \nu}{4} g_{\mu \nu} \Phi^6-\frac{\alpha}{2} (\nabla_\mu \Phi)(\nabla_\nu \Phi) \\
&+\frac{\beta}{8} g_{\mu \nu}\Phi^{-2} F^2_{\alpha \beta}+\frac{\beta}{2} \Phi^{-2} F_{\mu \alpha} F^\alpha{_\nu} +k C_{\mu \nu}=0.
\label{jkhl1}
\end{aligned}
\end{equation}
The variation of (\ref{witmg2}) with respect to $ A^\mu $ yields  
\begin{equation}
 (4 \sigma+\frac{\alpha}{4}) \nabla_\mu \Phi^2-(4 \sigma+\frac{\alpha}{4}) \Phi^2 A_\mu+\frac{k}{2} \epsilon_\mu{^{\lambda \nu}} \nabla_\lambda A_\nu - \beta \nabla^\nu (\Phi^{-2} F_{\mu\nu})=0. 
\end{equation}
And finally the variation of (\ref{witmg2}) with respect to $ \Phi $ results in
\begin{equation}
 2 \sigma \Phi \Big [R-4 \nabla .A-2 A^2 \Big]+\alpha \Big [\Box \Phi-\frac{1}{2} \Phi \nabla .A-\frac{1}{4} \Phi A^2-3\nu \Phi^5 \Big]+\frac{\beta}{2} \Phi^{-3} F^2_{\mu \nu}=0.
\end{equation}

\end{document}